\documentclass[12pt,preprint]{emulateapj}

\usepackage{natbib}
\usepackage{rotating}
\newcounter{column_number}
\begin{document}

\title{What is the Shell Around R Coronae Borealis?}

\author{ Edward J. Montiel\altaffilmark{1}, Geoffrey C. Clayton\altaffilmark{1}, Dominic C. Marcello\altaffilmark{1}, and Felix J. Lockman\altaffilmark{2}}
\altaffiltext{1}{Dept. of Physics \& Astronomy, Louisiana State University,
Baton Rouge, LA 70803; emonti2@lsu.edu, gclayton@fenway.phys.lsu.edu, dmarce1@tigers.lsu.edu}
\altaffiltext{2}{National Radio Astronomy Observatory, Green Bank, WV 24944; jlockman@nrao.edu}

\begin{abstract}
The hydrogen-deficient, carbon-rich R Coronae Borealis (RCB) stars are known for being prolific producers of dust which causes their large iconic declines in brightness. Several RCB stars, including R~CrB, itself, have large extended dust shells seen in the far-infrared. The origin of these shells is uncertain but they may give us clues to the evolution of the RCB stars. The shells could form in three possible ways. 1) they are fossil Planetary Nebula (PN) shells, which would exist if RCB stars are the result of a final, helium-shell flash, 2) they are material left over from a white-dwarf merger event which formed the RCB stars, or 3) they are material lost from the star during the RCB phase. Arecibo 21-cm observations establish an upper limit on the column density of H I in the R~CrB shell implying a maximum shell mass of $\lesssim$0.3 M$_{\sun}$. A low-mass fossil PN shell is still a possible source of the shell although it may not contain enough dust. The mass of gas lost during a white-dwarf merger event will not condense enough dust to produce the observed shell, assuming a reasonable gas-to-dust ratio. The third scenario where the shell around R~CrB has been produced during the star's RCB phase seems most likely to produce the observed mass of dust and the observed size of the shell. But this means that R CrB has been in its RCB phase for $\sim$10$^{4}$ yr.

\end{abstract}

%\keywords{dust, extinction --- ISM: abundances --- ISM: molecules --- ultraviolet: ISM} 

\section{Introduction}
 R Coronae Borealis (R CrB) is the prototype for its eponymous class of stars, which are very rare and have many unusual characteristics including extreme hydrogen deficiency and large, sudden declines in brightness of 8 magnitudes or more \citep{1994JApA...15...47L,1996PASP..108..225C,2012JAVSO..40..539C}. These declines are caused by clouds of carbon dust forming near the atmospheres of the stars, which are later dissipated by radiation pressure. Only about 100 R Coronae Borealis (RCB) stars are known in the Galaxy. Therefore, these stars may be the result of a rare form of stellar evolution or are in an evolutionary phase that lasts only a short time. 

Two scenarios have been suggested for producing an RCB star: the double degenerate (DD) and the final helium-shell flash (FF) models \citep{Iben:1996fj,2002MNRAS.333..121S}. In the DD model, an RCB star is the result of merger between a CO- and a He-white dwarf (WD) \citep{1984ApJ...277..355W}. The derivation of CNO isotopic ratios (in particular the excess of $^{18}$O) in RCB stars favors the DD scenario for these stars \citep{2005ApJ...623L.141C, 2007ApJ...662.1220C, 2009ApJ...696.1733G, 2010ApJ...714..144G}. In the FF model, a star evolving into a white dwarf undergoes a final helium-shell flash and expands to supergiant size \citep{Fujimoto:1977lr}. In this scenario, the star may have recently gone through a planetary nebula (PN) phase. Three stars (Sakurai's Object, V605 Aql, and FG Sge) have been observed to undergo FF outbursts that transformed them from hot evolved stars  into cool giants with spectroscopic properties similar to RCB stars \citep{1997AJ....114.2679C,1998ApJS..114..133G,1998A&A...332..651A,Asplund:1999bh,Asplund:2000qy,2006ApJ...646L..69C}. These FF stars are all surrounded by PNe. 

Several RCB stars have extended dust shells that are seen in reflected light in the visible or in emission from the far-IR \citep[e.g.,][]{Schaefer:1986lq,Walker:1985rr,1986ASSL..128..407W,2011MNRAS.414.1195B}. Infrared spectroscopy of RCB stars has been possible thanks to the Infrared Space Observatory (ISO) and the {\it Spitzer} Space Telescope, which has permitted the extraction of the characteristics of the IR emitting dust shell around RCBs (Lambert et al. 2001; Garc{\' i}a-Hern{\' a}ndez, Rao \& Lambert 2011a,b, 2013). It has been suggested that the mass loss in R CrB and V854 Cen is bipolar \citep{1993AJ....105.1915K,1997ApJ...476..870C}. This geometry is very common in PNe \citep{2002ARA&A..40..439B}. Further, the nebulosity, including cometary knots, seen around R~CrB and UW Cen has similar morphology to a PN shell \citep[][Clayton et al. 2015, in preparation]{2011ApJ...743...44C}. In this paper, we investigate whether the shell around R CrB is consistent with the scenarios suggested for the evolution of the RCB stars. 

\section{Arecibo 21-cm Observations}
Radio observations containing the region around R~CrB have been obtained from the first data release (DR1\footnote{https://purcell.ssl.berkeley.edu/}) Galactic Arecibo L-band Feed Array H I (GALFA-HI) Survey \citep{2011ApJS..194...20P}. The GALFA-HI survey provides both high resolution (4\arcmin) and high sensitivity (typical rms $\sim$ 80 mK) due to their 305-m aperture and the installation of the Arecibo L-band Feed Array (ALFA) \citep{2011ApJS..194...20P}. Fully reduced ``narrow" and ``wide" band data cubes were retrieved from the survey. The former provided 0.18 km s$^{-1}$ resolution in the local standard of rest (LSR) in the range v$_{LSR}$ = $\pm$190 km s$^{-1}$, while the latter provided 0.74 km s$^{-1}$ resolution in the range v$_{LSR}$ = $\pm$750 km s$^{-1}$. A full description of the data acquisition, which includes both drift and basketweave scanning, and reductions for the GALFA-HI survey are presented by \citet{2011ApJS..194...20P} (see their \S 3 and \S 4, respectively).

\section{Origin of the R~CrB Dust Shell}

\subsection{The R~CrB Dust Shell}

The foreground extinction toward R~CrB is quite small, E(B-V)$\sim$0.035 mag, since it lies at high Galactic latitude (b$^{II}$=+51\degr) \citep{1998ApJ...500..525S,2011ApJ...737..103S}. At maximum light, R~CrB is V=5.8 mag and B-V =0.6 mag \citep{Lawson:1990fk}. Based on the absolute magnitude/effective temperature relationship found for the Large Magellanic Cloud RCB stars, the absolute magnitude of R~CrB is estimated to be M$_V$=--5  mag  \citep{Alcock:2001lr,Tisserand:2009fj}. For the analysis in this paper, we thus adopt a distance to R~CrB of 1.4 kpc. The large extended far-IR shell, with a radius of $\sim$10\arcmin, surrounding  R~CrB was discovered with {\it IRAS} and then studied further with the {\it Spitzer}~and {\it Herschel} telescopes (Gillett et al. 1986; Clayton et al. 2011; Garc{\' i}a-Hern{\' a}ndez, Rao \& Lambert 2011b, 2013). At the assumed distance of R CrB, the radius of the shell corresponds to 4 pc. Monte Carlo radiative transfer modeling of the R~CrB shell suggests that it contains 10$^{-2}$ M$_{\sun}$ of dust \citep{2011ApJ...743...44C}.

\subsection{A Planetary Nebula Shell?}

The PNe around the FF stars, V605 Aql, Sakurai's Object, and FG Sge are still ionized. The shell around R~CrB is not. There is a small subclass of RCB stars that are much hotter (T$_{eff}$= 15,000-20,000 K) than the typical RCB stars and are surrounded by PNe \citep{1991MNRAS.248P...1P,2002AJ....123.3387D}. They are not hot enough at present to ionize their surrounding PNe, but the nebulae have not had time to recombine. The shells around the cooler RCB stars could be old PNe that have recombined. There are some strong similarities between the morphology of the shells of UW Cen and R~CrB, and some PNe such as the Eskimo Nebula. Cometary features seen in PNe such as the Eskimo are similar to those seen in these two RCB stars \citep{2011ApJ...743...44C}.
The recombination time depends on the electron densities of the shells and could range from hundreds to thousands of years. If n$_e$=200 cm$^{-3}$, as assumed for Sakurai's Object, the recombination time for R~CrB's PN shell would be $\sim$380 yr \citep{1999MNRAS.304..127P}. 
The amount of time that R~CrB has been an RCB star is unknown but it is one of the first variable stars to be discovered \citep{1797RSPT...87..133P}, so we know that it has been in its RCB star phase for at least 200 yr.
 
The PNe seen around the FF stars, Sakurai's Object and V605 Aql, are thought to be $\sim$2$\times$ 10$^{4}$ yr old, both with expansion velocities of $\sim$30 km s$^{-1}$ and radii of 0.35 pc for V605 Aql and 0.7 pc for Sakurai's Object \citep{1992MNRAS.257P..33P,1996ApJ...472..711G,1999MNRAS.304..127P}.  A FF star may not be old enough to produce a PN shell of the size seen around R~CrB as it only takes a few years for the star to reach its RCB-like phase after the FF \citep{2001ApJ...554L..71H}. 

If the R CrB shell is, in fact, a PN shell then it should be hydrogen rich \citep{Gillett:1986cr}. No pointed 21-cm measurements of any RCB circumstellar shells have been published previously. A visual inspection of the narrow band GALFA-HI data cube through velocity space was performed to see if there were any obvious H I features. The search focused on $\pm$30 km s$^{-1}$ of the v$_{LSR}$ of R~CrB $\sim$ 37 km s$^{-1}$ (The barycentric radial velocity of R~CrB $\sim$ 22.3 km s$^{-1}$  \citep{Lawson:1997fj}). No obvious structure was discerned. 

Next, a 20\arcmin~x 20\arcmin~square region, roughly the size of the dust shell, centered on R~CrB was selected, as well as a background region of the same size as shown in Figure 1. The background reference spectrum were subtracted from the spectrum toward the R~CrB shell to determine if there is any H I emission that might be associated with the shell.  While fluctuations in the background H I limit the comparison in many directions, one reference position, shown in Figure 1, has an H I spectrum that is identical within the noise to the spectrum averaged over the shell at the relevant velocities, and the difference spectrum (Figure 2) can be used to limit the H I mass of R~CrB.

The rms noise ($\sigma_{\rm rms}$) in the 0.18 km s$^{-1}$ channels of Figure 2 is 27 mK.  Over a $\pm30$ km s$^{-1}$ interval, this gives a $3\sigma$ limit on column density, N$_H$ (atoms cm$^{-2}$) in the R~CrB shell of 4.9 $\times$ 10$^{17}$. This was derived using N$_H$ = $3.0 \times 1.82 \times 10^{18} \times \sigma_{\rm rms} \times {\rm dv} \times \sqrt{{\rm n_{ch}}}$, where dv is the 21cm channel spacing (0.18 km s$^{-1}$) and n$_{\rm ch}$ is the number of channels ($\sim$330) over the relevant velocity range \citep{1990ARA&A..28..215D}. The mass of any neutral hydrogen, in solar masses, is then less than M$_H$ = 6.9 $\times$10$^{-28}$ $\times$ N$_H$ $\times$ D$^2$ $\times$ $\Omega$, where D is the distance in pc (1400 pc), and $\Omega$ is the solid angle of the region in arcmin$^2$.  With our upper limit on N$_H$, the corresponding $3\sigma$ limit on H~I in the R~CrB shell is 0.3 M$_{\sun}$. Assuming that the mass loss is contained in a shell that has been moving outward at 30 km s$^{-1}$, the PN shell would take $\sim$10$^5$ yr to expand to r= 4 pc. A shell of that age would almost certainly have had time to recombine and become neutral.

The number density of H in this shell would be low, similar to that seen in the diffuse interstellar medium. Therefore, the fraction of H$_2$ will be near zero, and all of the H will be atomic and neutral \citep{2009ApJS..180..125R}. The 21-cm measurements presented here put a strong upper limit on the H mass in the R~CrB shell. A study of the gas and dust masses in the shell of the PN, NGC 6781, gives a total shell mass of 0.86 M$_{\sun}$ and a dust mass of  4 $\times$ 10$^{-3}$ M$_{\sun}$ and therefore a gas-to-dust ratio of about 215 \citep{2014A&A...565A..36U}. Assuming a gas-to-dust ratio of 200, the amount of dust in the R~CrB shell would be $\lesssim$10$^{-3}$ M$_{\sun}$. 

\begin{figure}
\figurenum{1} 
\begin{center}
\includegraphics[width=3in,angle=0]{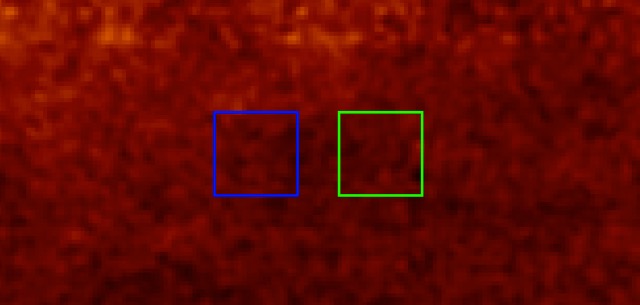}

\end{center}
\caption{GALFA-HI field (2\fdg25 x 1\fdg15) showing the location of the R~CrB shell (green square) and the area used for background subtraction (blue square). The squares are 20\arcmin~x 20\arcmin.}
\end{figure}

\begin{figure}
\figurenum{2} 
\begin{center}
\includegraphics[width=3in,angle=0]{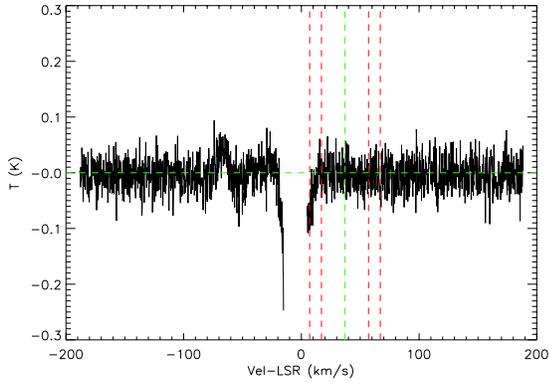}

\end{center}
\caption{A background-subtracted temperature vs. radial velocity plot for area around R~CrB. The vertical green dashed line is the position of R~CrB in radial velocity and the two vertical red lines mark $\pm$20 and $\pm$30 km s$^{-1}$ from the velocity of R CrB. The feature at -70 km s$^{-1}$ is part of a large elongated emission not associated with the R~CrB shell. The horizontal green dashed line is at temperature = 0\degr.}
\end{figure}

\subsection{Mass Loss from a White Dwarf Merger?}

We have simulated the RCB DD formation scenario by running three binary WD merger models using our fully three-dimensional adaptive mesh refinement (AMR) code (Marcello \& Tohline, in prep). This code evolves density, total energy, and angular, vertical, and radial momenta on a rotating Cartesian mesh. Once started, the model is driven into closer contact by artificially removing angular momentum from the system at a rate of $1\%$ per orbit for several orbits. Through AMR, we are able to run the models in large grids whose box dimensions are $20$--$30$ times larger than the orbital separation. This allows for a more accurate determination of mass loss from the system than on a smaller grid, where much of the material exiting the grid does not possess enough energy to escape. The initial conditions were created using the self-consistent field technique (SCF) to create a detached synchronously rotating binary WD in equilibrium just shy of becoming semi-detached \citep{1986ApJS...61..479H,2009ApJS..184..248E}. Each model run follows the merger of a CO- and a He-WD in a close binary orbit. The q=0.8 and q=0.5 models have a hybrid CO/He accretor, which as described by \citet{2012ApJ...757...76S} is a WD with a CO core and a He envelope of about 0.1 M$_{\sun}$. The results, along with the initial conditions, of our simulations are summarized in Table 1.

In the models presented here between $0.9$ and $3.3\%$ of the initial mass escapes and is not expected to fall back onto the new merged star. Therefore, in the three models described in Table 1 the mass lost from the two WDs, which could form into a circumstellar shell around the new RCB star, is $\sim$10$^{-2}$ M$_{\sun}$. The gas that has escape velocity is $>$95\% He so a maximum of 5\% of the gas could condense into dust. Assuming this and a different gas-to-dust ratio of 100 \citep{2003dge..conf.....W}, the amount of dust in such a shell would be $\sim$5 x 10$^{-6}$ M$_{\sun}$. A different gas-to-dust ratio was adopted, since in this scenario the envelope is not PN-like. The merger event itself only takes a few minutes after which there is a rapid expansion to supergiant size \citep{2012ApJ...757...76S}. \citet{2014MNRAS.445..660Z} estimate that the new star will expand to $\sim$200 R$_{\sun}$ in 500 yr. The estimated velocities of the gas escaping the grid in the simulations listed in Table 1 is 600--900 km s$^{-1}$. This gas would take $\sim$10$^4$ yr to expand to the size of the R~CrB shell. These velocities are the same order of magnitude as the winds measured in RCB stars in the He I $\lambda$10830 line \citep{2013AJ....146...23C}. This analysis assumes that any PN/common envelope phases of the two stars in the binary occurred long before the merger but this time period is not known. 

\begin{deluxetable}{llll}
%\rotate\
\tablecaption{Model Results}
\tablenum{1}
\scriptsize
%\begin{tabular}{ |c|c|c|c| }
\tablehead{\colhead{Parameter}&\colhead{Model 1}&\colhead{Model 2}&\colhead{Model 3}}
\startdata
%\hline
Mass Ratio (q)	 		&	$0.51 $				&	$0.70 $				&	$0.80	$	\\
Accretor Mass $(M_\odot)$	&	$0.503	$		&	$0.526$				&	$0.561	$	\\
Donor Mass ($M_\odot$)			&	$0.255	$		&	$0.366$				&	$0.45	$	\\
Total Mass ($M_\odot$)	 		&	$0.758	$		&	$0.892$				&	$1.011$		\\
\% Mass Lost	&	$2.3$				&	$0.90$			&	3.3		\\
Initial Period $(\mathrm{s})$		&	$181 $		&	$118$				&	$90.9$		\\
Final/Initial Period		&	$9.9 $	 			&	$2.9$				&	$6.9$		\\
Initial separation $(R_\odot)$		&	$6.30\times 10^{-2}$ 	&	$5.00\times 10^{-2}$		&	$4.38\times10^{-2}$	\\
Grid Box Dim. $(R_\odot)$		&	$1.74 $			&	$1$				&	$1.38$		\\
Orbits driven at 1\%		&	$2.5$				&	$4$				&	$2.5$		\\
dx $(R_\odot)$				& 	$5.66\times 10^{-4} $	&	$6.51\times 10^{-4}$	&	$4.49\times 10^{-4}$	\\
Init. Sep. in Cells	&	$111$				&	$77$				&	$98$ \\
%\hline
%\end{tabular}
\enddata
\end{deluxetable}

\begin{deluxetable}{lllll}
%\rotate\
\tablecaption{Shell Types}
\tablenum{2}
\scriptsize
%\begin{tabular}{ |c|c|c|c| }
\tablehead{\colhead{Type}&\colhead{v$_{exp}$}&\colhead{Time}&\colhead{Dust Mass}&\colhead{Total Mass}\\
\colhead{}&\colhead{(km s$^{-1}$)}&\colhead{(yr)}&\colhead{(M$_{\sun}$)}&\colhead{(M$_{\sun}$)}
}
\startdata
Planetary Nebula&30&10$^5$&$<$10$^{-3}$&$<$0.3\\
WD merger&600--900&10$^4$&5 x 10$^{-6}$&10$^{-2}$\\
RCB star&400&10$^4$&$>$10$^{-3}$&\nodata\\
\enddata
\end{deluxetable}

\vspace{0.7cm}

\subsection{RCB Phase Mass Loss?} 

It is thought that dust forms in puffs near the atmosphere of an RCB star \citep{1996PASP..108..225C}. We adopted the assumption that during a single dust formation event a puff forms at 2 R$_{\star}$ (R$_{\star}$=85 R$_{\sun}$) and subtends a fractional solid angle of 0.05, which results in the photosphere of the star being obscured and a dust mass of $\sim$10$^{-8}$ M$_{\sun}$ \citep{1992ApJ...397..652C,2011ApJ...743...44C}. Declines occur when a puff forms along the line of sight, but other puffs are likely forming around the star that do not cause declines. These puffs can be detected in the IR. Recent studies pertaining to IR variability of RCB stars has found that the covering factor can vary from RCB star to RCB star and find an average covering factor of 0.28$\pm$0.04 for R CrB (Garc{\' i}a-Hern{\' a}ndez, Rao \& Lambert 2011b; \citealt{2015MNRAS.447.3664R}). Other studies find a higher covering factor \citep{1984ApJ...280..228H,1999ApJ...517L.143C}. In addition, a correlation between pulsation phase and the timing of dust formation has been found in several RCB stars \citep{2007MNRAS.375..301C} and they typically show regular or semi-regular pulsation periods in the 40-100 d range \citep{Lawson:1990fk}. R~CrB, itself, does not have one regular period but has shown periods of 40 and 51 d \citep{1993MNRAS.265..899F}. Therefore, if a dust puff forms somewhere around R~CrB every 50 days then $\sim$10$^{-7}$ M$_{\sun}$ of dust will form per year around the star. 

There is strong evidence from observations at He I $\lambda$10830 that the dust, once formed, is accelerated quickly by radiation pressure from the star to $\sim$400 km s$^{-1}$ \citep{1992ApJ...397..652C,2003ApJ...595..412C,2013AJ....146...23C}. The He gas is likely dragged outward by the dust. Therefore, dust produced during the RCB phase could fill the R~CrB shell in only $\sim$10$^4$ yr. In that time, R~CrB will produce at least 10$^{-3}$ M$_{\sun}$ of dust depending how many puffs form around the star. Considering the assumptions made here and those made in the radiative transfer modeling, this estimated dust mass is close to the 10$^{-2}$ M$_{\sun}$ estimated by \citet{2011ApJ...743...44C}. The mass of gas in an RCB shell is unknown. The gas will be primarily He with little or no H. 

The observed far-IR shell around R CrB appears nearly spherical. The suggestion that the present day mass loss from R CrB is bipolar would not support this morphology \citep{1993AJ....105.1915K}. Further polarimetric or interferometric observations are needed to determine the morphology of the RCB-star dust mass loss \citep[e.g.,][]{1997ApJ...476..870C,2011MNRAS.414.1195B}.

\section{Summary}

The estimated masses of the circumstellar shells in the FF/PN, WD merger (DD), and RCB scenarios are summarized in Table 2. If the shell is an old fossil PN in the FF scenario, then the gas should be H-rich. In the other two cases, it would be H-poor and dominated by He gas. The results of the 21-cm observations find no detectable H in the R~CrB shell. The observations place upper limit of $\sim$0.3 M$_{\sun}$ on the mass of the PN. The ionized masses observed for some PNe  are in this range \citep[e.g.,][]{1994A&A...284..248B,1999MNRAS.308..623V}. The radius of R~CrB shell is very large for PN shell. The PN shells of the FF stars, V605 Aql and Sakurai's Object, are 0.35 and 0.7 pc, respectively.  A 4 pc radius PN shell would take $\sim$10$^5$ yr to fill at 30 km s$^{-1}$.  The gas mass loss in a WD merger is $\sim$10$^{-2}$ M$_{\sun}$. Since this gas is mostly He, little dust could form in such a shell. 

The gas mass of the R~CrB shell is not known, but the dust mass has been estimated to be 10$^{-2}$ M$_{\sun}$ \citep{2011ApJ...743...44C}. If it formed during the RCB phase then the shell would be filled with He gas, the mass of which cannot be measured so the gas-to-dust ratio is unknown. Based on the analyses above, we would expect only 10$^{-3}$ -- 10$^{-6}$ M$_{\sun}$ of dust in the PN and WD merger mass loss scenarios. Thus the suggestion that the R~CrB shell has formed from the dust forming during its present RCB phase seems most likely since it can form $\gtrsim$10$^{-3}$ M$_{\sun}$ of dust. If true, this model implies that R~CrB has been an RCB star for $\sim$10$^4$ yr to have produced a 4 pc radius shell. More sensitive 21-cm observations of RCB star shells are needed to place more stringent constraints on their H masses. 

We thank the anonymous referee for thoughtful suggestions that have improved this paper. This study was supported by NSF CREATIV grant AST-1240655. This study utilized data from Galactic ALFA HI (GALFA H I) survey data obtained with the 
Arecibo 305m telescope. The Arecibo Observatory is part of the National Astronomy and Ionosphere Center, which is operated by Cornell University under Cooperative Agreement with the NSF. The GALFA H I survey is funded by the NSF through grants to Columbia University, the University of Wisconsin, and the University of California.
The National Radio Astronomy Observatory is operated by Associated
Universities, Inc., under a cooperative agreement with the National
Science Foundation.
Portions of this research were conducted with high performance computational resources provided by Louisiana State University (http://www.hpc.lsu.edu).
This material is based upon work supported by the Louisiana Optical Network Institute (LONI).
This program also used the Extreme Science and Engineering Discovery Environment (XSEDE), which is supported by National Science Foundation grant number ACI-1053575.

%\bibliography{ms.bbl}

\begin{thebibliography}{50}
\expandafter\ifx\csname natexlab\endcsname\relax\def\natexlab#1{#1}\fi

\bibitem[{{Alcock} {et~al.}(2001){Alcock}, {Allsman}, {Alves}, {Axelrod},
  {Becker}, {Bennett}, {Clayton}, {Cook}, {Dalal}, {Drake}, {Freeman}, {Geha},
  {Gordon}, {Griest}, {Kilkenny}, {Lehner}, {Marshall}, {Minniti}, {Misselt},
  {Nelson}, {Peterson}, {Popowski}, {Pratt}, {Quinn}, {Stubbs}, {Sutherland},
  {Tomaney}, {Vandehei}, \& {Welch}}]{Alcock:2001lr}
{Alcock}, C., {et~al.} 2001, \apj, 554, 298

\bibitem[{{Asplund} {et~al.}(1998){Asplund}, {Gustafsson}, {Kameswara Rao}, \&
  {Lambert}}]{1998A&A...332..651A}
{Asplund}, M., {Gustafsson}, B., {Kameswara Rao}, N., \& {Lambert}, D.~L. 1998,
  \aap, 332, 651

\bibitem[{{Asplund} {et~al.}(2000){Asplund}, {Gustafsson}, {Lambert}, \&
  {Rao}}]{Asplund:2000qy}
{Asplund}, M., {Gustafsson}, B., {Lambert}, D.~L., \& {Rao}, N.~K. 2000, \aap,
  353, 287

\bibitem[{{Asplund} {et~al.}(1999){Asplund}, {Lambert}, {Kipper}, {Pollacco},
  \& {Shetrone}}]{Asplund:1999bh}
{Asplund}, M., {Lambert}, D.~L., {Kipper}, T., {Pollacco}, D., \& {Shetrone},
  M.~D. 1999, \aap, 343, 507
  
\bibitem[Balick 
\& Frank(2002)]{2002ARA&A..40..439B} Balick, B., \& Frank, A. 2002, \araa, 40, 439

\bibitem[{{Boffi} \& {Stanghellini}(1994)}]{1994A&A...284..248B}
{Boffi}, F.~R., \& {Stanghellini}, L. 1994, \aap, 284, 248

\bibitem[Bright et al.(2011)]{2011MNRAS.414.1195B} Bright, S.~N., Chesneau, 
O., Clayton, G.~C., et al. 2011, \mnras, 414, 1195 

\bibitem[{{Clayton} {et~al.}(1992){Clayton}, {Whitney}, {Stanford}, \&
  {Drilling}}]{1992ApJ...397..652C}
{Clayton}, G.~C., {Whitney}, B.~A., {Stanford}, S.~A., \& {Drilling}, J.~S.
  1992, \apj, 397, 652

\bibitem[{{Clayton}(1996)}]{1996PASP..108..225C}
{Clayton}, G.~C. 1996, \pasp, 108, 225

\bibitem[{{Clayton} \& {De Marco}(1997)}]{1997AJ....114.2679C}
{Clayton}, G.~C., \& {De Marco}, O. 1997, \aj, 114, 2679

\bibitem[Clayton et al.(1997)]{1997ApJ...476..870C} Clayton, G.~C., 
Bjorkman, K.~S., Nordsieck, K.~H., Zellner, N.~E.~B., 
\& Schulte-Ladbeck, R.~E. 1997, \apj, 476, 870 

\bibitem[{{Clayton} {et~al.}(1999){Clayton}, {Kerber}, {Gordon}, {Lawson},
  {Wolff}, {Pollacco}, \& {Furlan}}]{1999ApJ...517L.143C}
{Clayton}, G.~C., {Kerber}, F., {Gordon}, K.~D., {Lawson}, W.~A., {Wolff},
  M.~J., {Pollacco}, D.~L., \& {Furlan}, E. 1999, \apjl, 517, L143

\bibitem[{{Clayton} {et~al.}(2003){Clayton}, {Geballe}, \&
  {Bianchi}}]{2003ApJ...595..412C}
{Clayton}, G.~C., {Geballe}, T.~R., \& {Bianchi}, L. 2003, \apj, 595, 412

\bibitem[Clayton et al.(2005)]{2005ApJ...623L.141C} Clayton, G.~C., Herwig, 
F., Geballe, T.~R., et al. 2005, \apjl, 623, L141 

\bibitem[{{Clayton} {et~al.}(2006){Clayton}, {Kerber}, {Pirzkal}, {De Marco},
  {Crowther}, \& {Fedrow}}]{2006ApJ...646L..69C}
{Clayton}, G.~C., {Kerber}, F., {Pirzkal}, N., {De Marco}, O., {Crowther},
  P.~A., \& {Fedrow}, J.~M. 2006, \apjl, 646, L69

\bibitem[Clayton et al.(2007)]{2007ApJ...662.1220C} Clayton, G.~C., 
Geballe, T.~R., Herwig, F., Fryer, C., 
\& Asplund, M. 2007, \apj, 662, 1220 

\bibitem[{{Clayton} {et~al.}(2011){Clayton}, {Sugerman}, {Stanford}, {Whitney},
  {Honor}, {Babler}, {Barlow}, {Gordon}, {Andrews}, {Geballe}, {Bond}, {De
  Marco}, {Lawson}, {Sibthorpe}, {Olofsson}, {Polehampton}, {Gomez},
  {Matsuura}, {Hargrave}, {Ivison}, {Wesson}, {Leeks}, {Swinyard}, \&
  {Lim}}]{2011ApJ...743...44C}
{Clayton}, G.~C., {et~al.} 2011, \apj, 743, 44

\bibitem[{{Clayton}(2012)}]{2012JAVSO..40..539C}
---. 2012, JAAVSO, 40, 539

\bibitem[{{Clayton} {et~al.}(2013){Clayton}, {Geballe}, \&
  {Zhang}}]{2013AJ....146...23C}
{Clayton}, G.~C., {Geballe}, T.~R., \& {Zhang}, W. 2013, \aj, 146, 23

\bibitem[{{Crause} {et~al.}(2007){Crause}, {Lawson}, \&
  {Henden}}]{2007MNRAS.375..301C}
{Crause}, L.~A., {Lawson}, W.~A., \& {Henden}, A.~A. 2007, \mnras, 375, 301

\bibitem[{{De Marco} {et~al.}(2002){De Marco}, {Clayton}, {Herwig}, {Pollacco},
  {Clark}, \& {Kilkenny}}]{2002AJ....123.3387D}
{De Marco}, O., {Clayton}, G.~C., {Herwig}, F., {Pollacco}, D.~L., {Clark},
  J.~S., \& {Kilkenny}, D. 2002, \aj, 123, 3387

%\bibitem[{{Dehnen}(2000)}]{2000ApJ...536L..39D}{Dehnen}, W. 2000, \apjl, 536, L39

\bibitem[Dickey \& Lockman(1990)]{1990ARA&A..28..215D} Dickey, J.~M., \& Lockman, F.~J. 1990, \araa, 28, 215

\bibitem[{{Even} \& {Tohline}(2009)}]{2009ApJS..184..248E}
{Even}, W., \& {Tohline}, J.~E. 2009, \apjs, 184, 248

\bibitem[{{Fernie} \& {Lawson}(1993)}]{1993MNRAS.265..899F}
{Fernie}, J.~D., \& {Lawson}, W.~A. 1993, \mnras, 265, 899

\bibitem[{{Fujimoto}(1977)}]{Fujimoto:1977lr}
{Fujimoto}, M.~Y. 1977, \pasj, 29, 331

\bibitem[Garc{\'{\i}}a-Hern{\'a}ndez et al.(2009)]{2009ApJ...696.1733G} 
Garc{\'{\i}}a-Hern{\'a}ndez, D.~A., Hinkle, K.~H., Lambert, D.~L., 
\& Eriksson, K. 2009, \apj, 696, 1733 

\bibitem[Garc{\'{\i}}a-Hern{\'a}ndez et al.(2010)]{2010ApJ...714..144G} 
Garc{\'{\i}}a-Hern{\'a}ndez, D.~A., Lambert, D.~L., Kameswara Rao, N., 
Hinkle, K.~H., \& Eriksson, K. 2010, \apj, 714, 144

\bibitem[Garc{\'{\i}}a-Hern{\'a}ndez et al.(2011a)]{2011ApJ...729..126G}  
Garc{\'{\i}}a-Hern{\'a}ndez, D.~A., Kameswara Rao, N., 
\& Lambert, D.~L. 2011a, \apj, 729, 126 

\bibitem[{{Garc{\'{\i}}a-Hern{\'a}ndez}
  {et~al.}(2011b){Garc{\'{\i}}a-Hern{\'a}Ned}, {Kameswara Rao}, \&
  {Lambert}}]{2011ApJ...739...37G}
{Garc{\'{\i}}a-Hern{\'a}ndez}, D.~A., {Kameswara Rao}, N., \& {Lambert}, D.~L.
  2011b, \apj, 739, 37

\bibitem[Garc{\'{\i}}a-Hern{\'a}ndez et al.(2013)]{2013ApJ...773..107G} 
Garc{\'{\i}}a-Hern{\'a}ndez, D.~A., Kameswara Rao, N., 
\& Lambert, D.~L. 2013, \apj, 773, 107 

\bibitem[{{Gillett} {et~al.}(1986){Gillett}, {Backman}, {Beichman}, \&
  {Neugebauer}}]{Gillett:1986cr}
{Gillett}, F.~C., {Backman}, D.~E., {Beichman}, C., \& {Neugebauer}, G. 1986,
  \apj, 310, 842

\bibitem[{{Gonzalez} {et~al.}(1998){Gonzalez}, {Lambert}, {Wallerstein}, {Rao},
  {Smith}, \& {McCarthy}}]{1998ApJS..114..133G}
{Gonzalez}, G., {Lambert}, D.~L., {Wallerstein}, G., {Rao}, N.~K., {Smith},
  V.~V., \& {McCarthy}, J.~K. 1998, \apjs, 114, 133

\bibitem[{{Guerrero} \& {Manchado}(1996)}]{1996ApJ...472..711G}
{Guerrero}, M.~A., \& {Manchado}, A. 1996, \apj, 472, 711

\bibitem[{{Hachisu}(1986)}]{1986ApJS...61..479H}{Hachisu}, I. 1986, \apjs, 61, 479

\bibitem[Hecht et al.(1984)]{1984ApJ...280..228H} Hecht, J.~H., Holm, 
A.~V., Donn, B., \& Wu, C.-C.\ 1984, \apj, 280, 228

\bibitem[{{Herwig}(2001)}]{2001ApJ...554L..71H}
{Herwig}, F. 2001, \apjl, 554, L71

\bibitem[{{Iben} {et~al.}(1996){Iben}, {Tutukov}, \& {Yungelson}}]{Iben:1996fj}
{Iben}, Jr., I., {Tutukov}, A.~V., \& {Yungelson}, L.~R. 1996, \apj, 456, 750

\bibitem[Lambert \& Rao(1994)]{1994JApA...15...47L} Lambert, D.~L., \& Rao, N.~K. 1994, Journal of Astrophysics and Astronomy, 15, 47 

\bibitem[{{Lambert} {et~al.}(2001){Lambert}, {Rao}, {Pandey}, \&
  {Ivans}}]{2001ApJ...555..925L}
{Lambert}, D.~L., {Rao}, N.~K., {Pandey}, G., \& {Ivans}, I.~I. 2001, \apj,
  555, 925

\bibitem[{{Lawson} \& {Cottrell}(1997)}]{Lawson:1997fj}
{Lawson}, W.~A., \& {Cottrell}, P.~L. 1997, \mnras, 285, 266

\bibitem[{{Lawson} {et~al.}(1990){Lawson}, {Cottrell}, {Kilmartin}, \&
  {Gilmore}}]{Lawson:1990fk}
{Lawson}, W.~A., {Cottrell}, P.~L., {Kilmartin}, P.~M., \& {Gilmore}, A.~C.
  1990, \mnras, 247, 91

\bibitem[{{Peek} {et~al.}(2011){Peek}, {Heiles}, {Douglas}, {Lee}, {Grcevich},
  {Stanimirovi{\'c}}, {Putman}, {Korpela}, {Gibson}, {Begum}, {Saul},
  {Robishaw}, \& {Kr{\v c}o}}]{2011ApJS..194...20P}
{Peek}, J.~E.~G., {et~al.} 2011, \apjs, 194, 20

\bibitem[{{Pigott} \& {Englefield}(1797)}]{1797RSPT...87..133P}
{Pigott}, E., \& {Englefield}, H.~C. 1797, Royal Society of London
  Philosophical Transactions Series I, 87, 133

\bibitem[{{Pollacco}(1999)}]{1999MNRAS.304..127P}
{Pollacco}, D. 1999, \mnras, 304, 127

\bibitem[{{Pollacco} {et~al.}(1991){Pollacco}, {Hill}, {Houziaux}, \&
  {Manfroid}}]{1991MNRAS.248P...1P}
{Pollacco}, D.~L., {Hill}, P.~W., {Houziaux}, L., \& {Manfroid}, J. 1991,
  \mnras, 248, 1P

\bibitem[{{Pollacco} {et~al.}(1992){Pollacco}, {Lawson}, {Clegg}, \&
  {Hill}}]{1992MNRAS.257P..33P}
{Pollacco}, D.~L., {Lawson}, W.~A., {Clegg}, R.~E.~S., \& {Hill}, P.~W. 1992,
  \mnras, 257, 33P

\bibitem[{{Rachford} {et~al.}(2009){Rachford}, {Snow}, {Destree}, {Ross},
  {Ferlet}, {Friedman}, {Gry}, {Jenkins}, {Morton}, {Savage}, {Shull},
  {Sonnentrucker}, {Tumlinson}, {Vidal-Madjar}, {Welty}, \&
  {York}}]{2009ApJS..180..125R}
{Rachford}, B.~L., {et~al.} 2009, \apjs, 180, 125

\bibitem[Rao \& Lambert(1993)]{1993AJ....105.1915K} Rao, N.,~K. \& Lambert, D.~L. 1993, \aj, 105, 1915

\bibitem[Rao \& Lambert(2015)]{2015MNRAS.447.3664R} Rao, N.~K., \& Lambert, D.~L. 2015, \mnras, 447, 3664

\bibitem[{{Saio} \& {Jeffery}(2002)}]{2002MNRAS.333..121S}
{Saio}, H., \& {Jeffery}, C.~S. 2002, \mnras, 333, 121

\bibitem[{{Schaefer}(1986)}]{Schaefer:1986lq}
{Schaefer}, B.~E. 1986, \apj, 307, 644

\bibitem[{{Schlafly} \& {Finkbeiner}(2011)}]{2011ApJ...737..103S}
{Schlafly}, E.~F., \& {Finkbeiner}, D.~P. 2011, \apj, 737, 103

\bibitem[{{Schlegel} {et~al.}(1998){Schlegel}, {Finkbeiner}, \&
  {Davis}}]{1998ApJ...500..525S}
{Schlegel}, D.~J., {Finkbeiner}, D.~P., \& {Davis}, M. 1998, \apj, 500, 525

\bibitem[{{Staff} {et~al.}(2012){Staff}, {Menon}, {Herwig}, {Even}, {Fryer},
  {Motl}, {Geballe}, {Pignatari}, {Clayton}, \&
  {Tohline}}]{2012ApJ...757...76S}
{Staff}, J.~E., {et~al.} 2012, \apj, 757, 76

\bibitem[{{Tisserand} {et~al.}(2009){Tisserand}, {Wood}, {Marquette}, {Afonso},
  {Albert}, {Andersen}, {Ansari}, {Aubourg}, {Bareyre}, {Beaulieu}, {Charlot},
  {Coutures}, {Ferlet}, {Fouqu{\'e}}, {Glicenstein}, {Goldman}, {Gould},
  {Gros}, {de Kat}, {Lesquoy}, {Loup}, {Magneville}, {Maurice}, {Maury},
  {Milsztajn}, {Moniez}, {Palanque-Delabrouille}, {Perdereau}, {Rich},
  {Schwemling}, {Spiro}, \& {Vidal-Madjar}}]{Tisserand:2009fj}
{Tisserand}, P., {et~al.} 2009, \aap, 501, 985

\bibitem[{{Ueta} {et~al.}(2014){Ueta}, {Ladjal}, {Exter}, {Otsuka}, {Szczerba},
  {Si{\'o}dmiak}, {Aleman}, {van Hoof}, {Kastner}, {Montez}, {McDonald},
  {Wittkowski}, {Sandin}, {Ramstedt}, {De Marco}, {Villaver}, {Chu},
  {Vlemmings}, {Izumiura}, {Sahai}, {Lopez}, {Balick}, {Zijlstra}, {Tielens},
  {Rattray}, {Behar}, {Blackman}, {Hebden}, {Hora}, {Murakawa}, {Nordhaus},
  {Nordon}, \& {Yamamura}}]{2014A&A...565A..36U}
{Ueta}, T., {et~al.} 2014, \aap, 565, A36

\bibitem[{{van Hoof} \& {van de Steene}(1999)}]{1999MNRAS.308..623V}
{van Hoof}, P.~A.~M., \& {van de Steene}, G.~C. 1999, \mnras, 308, 623

\bibitem[{{Walker}(1985)}]{Walker:1985rr}
{Walker}, H.~J. 1985, \aap, 152, 58

\bibitem[{{Walker}(1986)}]{1986ASSL..128..407W}
{Walker}, H.~J. 1986, in Astrophysics and Space Science Library, Vol. 128, IAU
  Colloq. 87: Hydrogen Deficient Stars and Related Objects, ed. K.~{Hunger},
  D.~{Schoenberner}, \& N.~{Kameswara Rao}, 407

\bibitem[{{Webbink}(1984)}]{1984ApJ...277..355W}
{Webbink}, R.~F. 1984, \apj, 277, 355

\bibitem[{{Whittet}(2003)}]{2003dge..conf.....W}
{Whittet}, D.~C.~B. 2003, {Dust in the galactic environment}, 2nd edn.
  (Bristol: Institute of Physics (IOP) Publishing)

\bibitem[{{Zhang} {et~al.}(2014){Zhang}, {Jeffery}, {Chen}, \&
  {Han}}]{2014MNRAS.445..660Z}
{Zhang}, X., {Jeffery}, C.~S., {Chen}, X., \& {Han}, Z. 2014, \mnras, 445, 660

\end{thebibliography}

\end{document}